\documentclass[aps,pra,amsfonts,amsmath,amssymb,reprint,superscriptaddress,floatfix]{revtex4-2}
\usepackage{graphicx, hyperref}

\newcommand*{\avg}[1]{\langle #1 \rangle} % for average
\newcommand*{\abs}[1]{\left| #1 \right|} % for absolute value
\newcommand{\ket}[1]{\left| #1 \right>} % for Dirac bras
\newcommand{\bra}[1]{\left< #1 \right|} % for Dirac kets
\newcommand{\braket}[2]{\left< #1 \vphantom{#2} | #2 \vphantom{#1} \right>} % for Dirac brackets
\DeclareMathOperator{\Tr}{Tr}
\DeclareMathOperator{\var}{var}
\DeclareMathOperator{\real}{Re}
\begin{document}

\title{Saddle-point scrambling without thermalisation}
\author{R. A. Kidd}
\email{ryan.kidd@uq.edu.au}
\affiliation{School of Mathematics and Physics, University of Queensland, Brisbane, Queensland 4072, Australia.}
\author{A. Safavi-Naini}
\affiliation{ARC Centre of Excellence for Engineered Quantum Systems,
School of Mathematics and Physics, University of Queensland, Brisbane, Queensland 4072, Australia}
\affiliation{School of Mathematics and Physics, University of Queensland, Brisbane, Queensland 4072, Australia.}
\affiliation{Institute for Theoretical Physics, University of Amsterdam, Science Park 904, 1098 XH Amsterdam, the Netherlands}
\author{J. F. Corney}
\affiliation{School of Mathematics and Physics, University of Queensland, Brisbane, Queensland 4072, Australia.}

%-----------------------------------------------------------------------
\date{\today}
%------------------------------------------------------------------------

\begin{abstract}

Out-of-time-order correlators (OTOCs) have proven to be a useful tool for studying thermalisation in quantum systems. In particular, the exponential growth of OTOCS, or scrambling, is sometimes taken as an indicator of chaos in quantum systems, despite the fact that saddle points in integrable systems can also drive rapid growth in OTOCs. By analysing the Dicke model and a driven Bose-Hubbard dimer, we demonstrate that the OTOC growth driven by chaos can, nonetheless, be distinguished from that driven by saddle points through the long-term behaviour. Besides quantitative differences in the long-term average, the saddle point gives rise to large oscillations not observed in the chaotic case. The differences are also highlighted by entanglement entropy, which in the chaotic driven dimer matches a Page curve prediction. These results illustrate additional markers that can be used to distinguish chaotic behaviour in quantum systems, beyond the initial exponential growth in OTOCs.

\end{abstract}


%*******************************************

\maketitle

\section{Introduction}

The dynamics of quantum information provides a link among different areas of physics, as is evident in the study of fast scrambling, quantified by the exponential growth of the out-of-time-order correlators (OTOCs)~\cite{Larkin1969, *Larkin1969_2, Hayden2007, Sekino2008, Shenker2014, Roberts2015, Hosur2016, Maldacena2016}. First introduced to describe the dynamics of quantum information in black holes, scrambling has since been used to probe the connections between the dynamics of entanglement, chaos, and thermalisation~\cite{Rozenbaum2017, Rammensee2018, Garcia-Mata2018, Lewis-Swan2019, Prakash2020, Xu2020}. In particular, studies of chaotic models in periodically driven and undriven systems have used a variety of OTOCs, including the fidelity out-of-time-order correlator (FOTOC), to show that, in the chaotic phase, the OTOC grows, often exponentially, up to an Ehrenfest time, after which it saturates to a steady state value~\cite{Garttner2017, Garttner2018, Lewis-Swan2019, Pilatowsky-Cameo2020, Kidd2020, Rozenbaum2017, Rammensee2018, Garcia-Mata2018, Fortes2019, Fortes2020, Prakash2020, Kidd2020}. If the state of an isolated quantum system is sufficiently delocalised in the basis of energy eigenstates, the system is expected to relax towards the `diagonal ensemble' (DE) due to dephasing between energy eigenstates~\cite{DAlessio2016}. Consequently, in the absence of an external drive an initial product state will evolve to a finite-temperature thermal state with volume-law entanglement entropy, whereas a periodically driven system will evolve to an infinite-temperature state~\cite{Garttner2017, Garttner2018, Lewis-Swan2019, Kidd2020, DAlessio2016}.

Although the exponential growth of OTOCs is often associated with chaos, it has been demonstrated recently that, in the absence of chaos, the exponential growth of an OTOC may be driven by an unstable trajectory associated with a hyperbolic fixed point---or saddle point---in the semiclassical phase space~\cite{Pappalardi2018, Hummel2019, Bhattacharyya2019, Ali2020, Pilatowsky-Cameo2020, Xu2020, Hashimoto2020, Bhattacharyya2021}.  So far, scrambling without chaos has been observed in Ising spin chains with long-range interactions~\cite{Pappalardi2018}, the truncated Lieb-Liniger model~\cite{Hummel2019}, the inverted harmonic oscillator~\cite{Bhattacharyya2019, Ali2020, Hashimoto2020, Bhattacharyya2021}, the Dicke model~\cite{Pilatowsky-Cameo2020} and the Lipkin-Meshkov-Glick (LMG) model~\cite{Xu2020}. Saddle-dominated scrambling in the presence of chaos has been observed in the kicked rotor model, the Feingold-Peres model of coupled tops and elastic manifolds pinned in a random potential~\cite{Xu2020}.

In this paper we study the entanglement dynamics in the two-site Bose-Hubbard model, showing explicit differences between the chaotic and the saddle point regimes. This model, which could be experimentally implemented in ultracold atoms, allows access to stable, unstable (chaotic), and saddle-point regions in phase-space through tuning and modulation of the inter-well tunnelling rate. We solve exactly the short-time and long-time (times much longer than the Ehrenfest time) dynamics for up to $10^3$ particles. In the chaotic regime, OTOCs grow exponentially until saturating at the infinite temperature prediction, consistent with Floquet thermalisation. In contrast, we find that the saddle point FOTOC exhibits large, long-lived oscillations around the diagonal ensemble prediction. These differences can be traced back to distinct differences in the respective diagonal ensembles, reflected quantitatively in the Shannon entropy. We also study the long-time behaviour of the von Neumann entanglement entropy, and its scaling with subsystem size, to highlight the differences between the two regimes and confirm the lack of thermalisation at the saddle point.

To demonstrate that these ideas generalise to higher-dimensional phase spaces, we also calculate FOTOC behaviour for the well-known Dicke model, which comprises noninteracting spins coupled to a photon mode~\cite{Dicke1954}.  By varying the relative strength of the spin-photon coupling, one can tune to chaotic regimes without the need for external driving.  Here we find short-term scrambling in the presence of the saddle-point, regardless of the extent of the semiclassical chaos.  However, it is again only in the chaos-dominated regime where long-time thermal behaviour is evident.

\begin{figure*}[t]
    \centering
    \includegraphics[width=\textwidth]{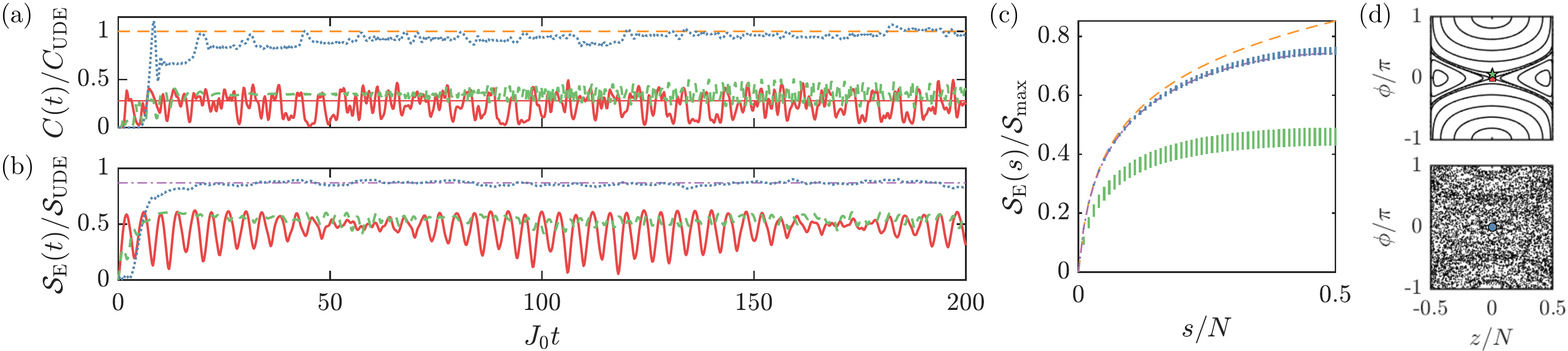}
    \caption{\label{fig:FOTOC}a) FOTOCs for driven-dimer chaotic phase space (blue dotted line), a saddle point (thick red line) and a point slightly perturbed from the saddle point (green dashed line) for $N=1000$ particles. The infinite-temperature uniform diagonal ensemble (UDE) prediction is indicated by the horizontal thin orange dashed line. The saddle point state diagonal ensemble prediction is indicated by the horizontal thin red line.
    b) Driven dimer von Neumann entanglement entropy for a subsystem of $s=N/2$ out of a total of $N=100$ particles. The Page curve prediction is indicated by the thin purple dash-dotted line.
    c) Post-Ehrenfest time-averaged von Neumann entanglement entropy for the driven-dimer chaotic (blue error bars) and saddle-perturbed (green error bars) states, with two standard deviations indicated by bar height. The Page curve prediction (purple dashed-dot line) coincides with the chaotic state result (blue error bars). Entanglement entropy is averaged over $20 \leq J_0t \leq 200$.
    d) Driven dimer semiclassical phase space for the regimes with a saddle point (top: $NU = -2$, $J=1$) and chaos (bottom: $NU=-1$, $J(t) = 1+1.5\cos{(0.5t)}$). The perturbed state $(z,\phi/\pi) = (0,0.06)$ is indicated by the green star and the saddle point and chaotic states $(z,\phi/\pi) = (0,0)$ are indicated by the red square and blue circle, respectively. The phase space representation is periodic in $\phi \in [-\pi,\pi)$.}
\end{figure*}

Although much previous work has focused on the short-term dynamics of OTOCs, other authors have also recently turned to the long-term dynamics to provide a less ambiguous probe of chaos in quantum systems.  Apart from the issues associated with saddle points that we focus on here, the short-term dynamics can be ambiguous in a mixed phase space, with regular and chaotic regions, especially when delocalised initial states are used with the OTOCs.  In such cases, the long term behaviour provides a clearer indication of the transition between integrability and chaos, as seen for example in the quantum kicked rotor~\cite{Rozenbaum2017}, Bose-Hubbard systems~\cite{Rammensee2018}, quantum maps~\cite{Garcia-Mata2018, Fortes2019} and spin chains~\cite{Fortes2019, Fortes2020}.

\section{Models}

In this paper we primarily consider the Bose-Hubbard dimer~\cite{Milburn1997}, which is described by the Hamiltonian,
\begin{equation} \label{eq:H}
    \hat{H}_{B} = 2\hbar U \hat{S}_z^2 - 2\hbar J\hat{S}_x,
  \end{equation}
where $J$ is the tunneling rate and $U$ is the on-site interaction strength. We use the pseudo-angular-momentum operators $\hat{S}_\alpha$ with $\alpha= x, y, z$,
\begin{eqnarray}
    \hat{S}_x &=& \frac{\hat{a}_2^\dagger \hat{a}_1 + \hat{a}_1^\dagger \hat{a}_2}{2}, \nonumber \\
    \hat{S}_y &=& \frac{\hat{a}_2^\dagger \hat{a}_1 - \hat{a}_1^\dagger \hat{a}_2}{2i}, \nonumber \\
    \hat{S}_z &=& \frac{\hat{a}_2^\dagger \hat{a}_2 - \hat{a}_1^\dagger \hat{a}_1}{2},
\end{eqnarray}
where $\hat{a}_j$, $\hat{a}_j^\dagger$ with $j=1,2$ are the creation and annihilation operators for the two bosonic modes and $[\hat{S}_\alpha, \hat{S}_\beta] = i \epsilon_{\alpha \beta \gamma} \hat{S}_\gamma$.  The Bose-Hubbard dimer can be mapped to a particular instance of the LMG model, whose saddle-point OTOC dynamics was studied in Refs.~\cite{Xu2020, Pilatowsky-Cameo2020}.

The semiclassical phase space of the two-site Bose-Hubbard model is shown in Fig.~\ref{fig:FOTOC}(d) in two different regimes, using coordinates $z=\avg{\hat{S}_z}$ and $\phi=-\arg{( \avg{\hat{S}_x}+i\avg{\hat{S}_y} )}$. For $\left|NU/J\right|\leq 1$, the system undergoes Rabi oscillations and the fixed points are two stable centres~\cite{Milburn1997}. At the critical interaction strength, $\left|NU/J\right| = 1$, one of the stable centres undergoes a pitchfork bifurcation.  For stronger interactions, `self-trapping trajectories' emerge on either side of a hyperbolic fixed point~\cite{Milburn1997}, as shown in the upper plot in Fig.~\ref{fig:FOTOC}(d).

The addition of periodic modulation to the tunneling frequency, $J(t) = J_0[1 + \mu \cos{(\omega t)}]$ makes the two-site Bose-Hubbard model chaotic~\cite{Milburn1997chaos}, as shown in the lower plot in Fig.~\ref{fig:FOTOC}(d). The extent of the chaos can be finely controlled through modulation of the constants $\mu,\omega$~\cite{Milburn1997chaos, Kidd2019, Kidd2020}.

\begin{figure}[ht]
    \centering
    \includegraphics[width=\columnwidth]{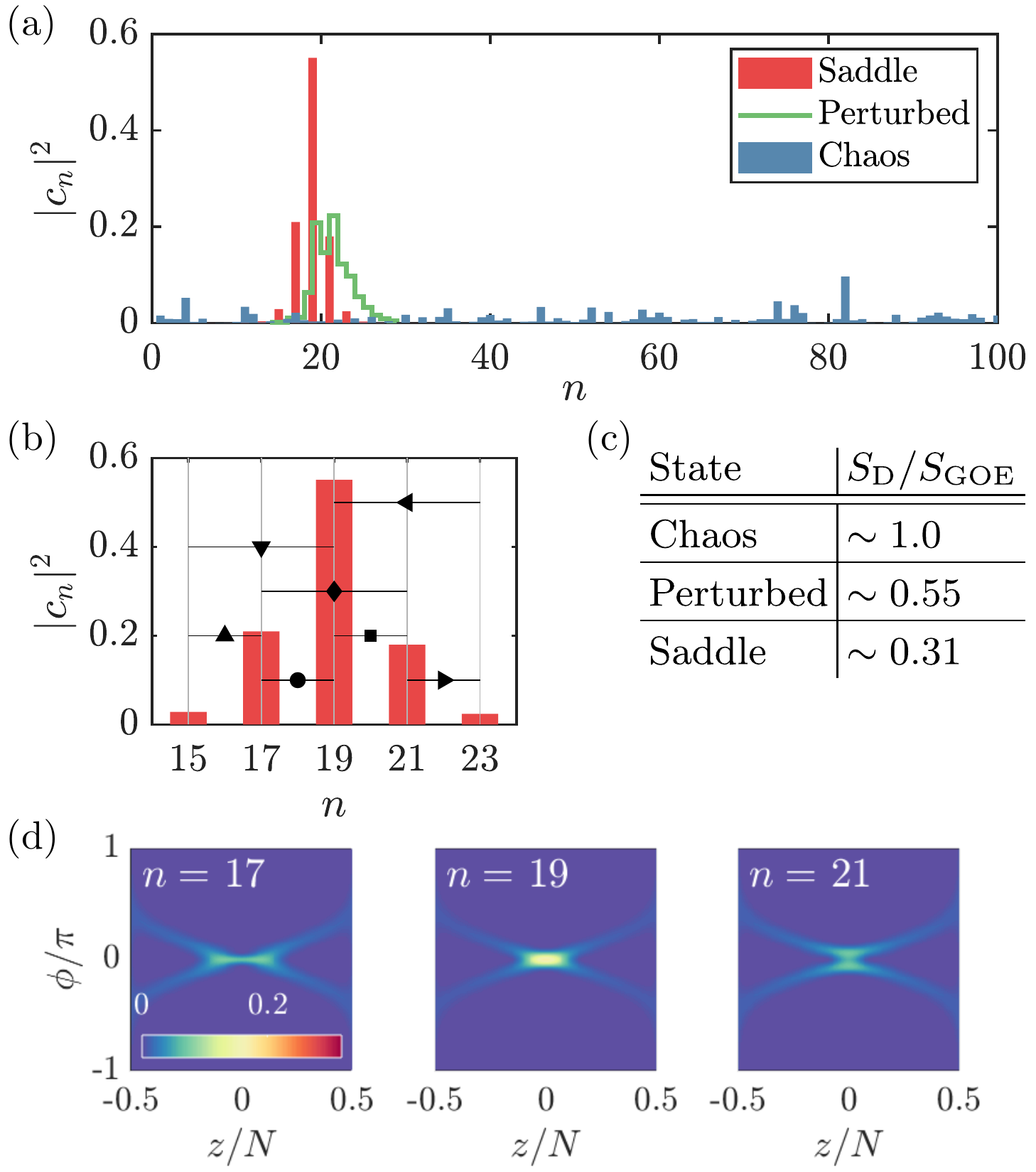}
    \caption{\label{fig:diag_ensembles}
    a) Diagonal distributions of coherent states in the driven-dimer chaotic phase space (blue bars), on the saddle point (red bars) and slightly perturbed from the saddle point (green staircase), as in Fig.~\ref{fig:FOTOC}. The distributions are ordered by energy or quasienergy magnitude and indexed by $n$.
    b) Close-up of driven-dimer saddle-point diagonal distribution with selected high-overlap energy eigenstate transitions labelled by black markers, corresponding to frequencies identified in Figs.~\ref{fig:FOTOC_fft}(a) and \ref{fig:FOTOC_fft}(b).
    c) Diagonal entropies of the initial coherent states used in the driven dimer simulations, relative to the Gaussian orthogonal ensemble (GOE) prediction.
    d) Driven-dimer saddle-point eigenstate $Q$ distributions for $N=100$, labelled by $n$. The three $Q$ distributions shown are those with the largest overlap with the saddle-point coherent state. Parameters are as in Fig.~\ref{fig:FOTOC}.}
\end{figure}

\begin{figure*}[t]
    \centering
    \includegraphics[width=\textwidth]{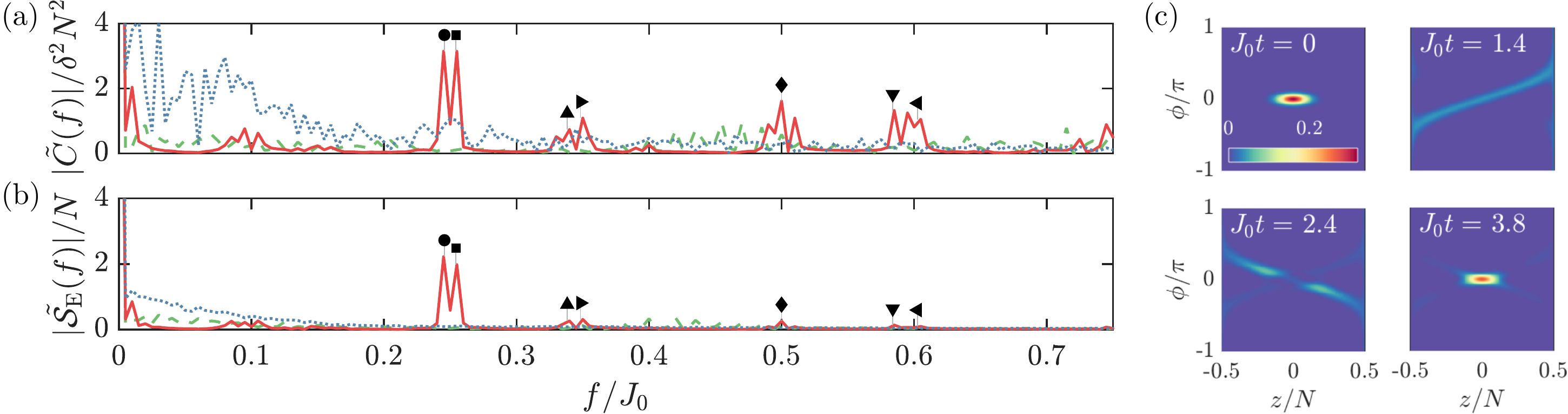}
    \caption{\label{fig:FOTOC_fft}a) Fourier spectra of FOTOCs for driven-dimer chaotic phase space (blue dotted line), a saddle point (red line) and a point slightly perturbed from the saddle point (green dashed line) for $N=100$ particles.
    b) Fourier spectra of driven-dimer von Neumann entanglement entropy for a subsystem of $s=N/2$ particles from a total of $N=100$ particles.
    In both plots, high-occupancy energy eigenstate transition frequencies for the saddle point state are indicated by black markers, corresponding to Fig.~\ref{fig:diag_ensembles}(b).
    Parameters are as in Fig.~\ref{fig:FOTOC}.
    c) Driven-dimer saddle-point state $Q$ distribution dynamics for $N=100$.}
\end{figure*}

To consider systems with higher dimensional phase space, we employ the Dicke model, comprising noninteracting spins coupled to a photon mode~\cite{Dicke1954}. Describing the spins collectively through the use of the pseudo angular momentum operators, $\hat{S}_\alpha$, and introducing photon annihilation (creation) operators, $\hat{b}$ ($\hat{b}^\dagger$), one can write the Dicke model Hamiltonian as
\begin{equation}
    \hat{H}_D = \hbar \omega \hat{b}^\dagger \hat{b} + \hbar \Delta \hat{S}_z + \frac{2\hbar\gamma}{\sqrt{N}} (\hat{b}^\dagger + \hat{b}) \hat{S}_x,
\end{equation}
where $\omega$ is the optical frequency, $\Delta$ is the atomic transition frequency, $\gamma$ is the atom-field coupling, and $N$ is the number of spins. The semiclassical version of the Dicke model is integrable only when one of the three parameters $\omega$, $\Delta$ and $\gamma$ is zero, and it exhibits widespread chaos in its four-dimensional phase space at high energies~\cite{Chavez-Carlos2016, Bastarrachea-Magnani2017, Chavez-Carlos2019}.

For atom-field coupling at or below the critical value of $\gamma_c = \sqrt{\omega\Delta}/2$, the Dicke-model ground state comprises the spin-down Bloch coherent state of the atomic subsystem and the vacuum coherent state of the photonic subsystem. Above the critical value, the spin-down-vacuum state is no longer the Dicke ground state, and chaos is widespread in its phase space energy shell~\cite{Chavez-Carlos2016, Bastarrachea-Magnani2016, Bastarrachea-Magnani2017, Chavez-Carlos2019}. The spin-down-vacuum state corresponds to a saddle point in the semiclassical phase space and will hereafter be referred to as the saddle-point state of the Dicke model.

As the Dicke model is nonintegrable in the regimes we study and has an infinite-dimensional Hilbert space due to the photon subspace, we diagonalise the Hamiltonian in a truncated basis. We solve the Dicke model in the so-called `efficient coherent basis'~\cite{ECB0, ECB1, ECB2}, which allows calculation of converged energy eigenvalues for a relatively small number of basis modes. We ensure convergence of energy eigenstates by comparison with the next smallest truncated basis and only retain eigenstates that differ in energy by less than $10^{-3}$. The basis truncation is chosen such that the normalisation of the initial state expressed in the energy eigenbasis differs from unity by no more than $10^{-5}$.

\section{Results}

\subsection{FOTOCs}

To study the entanglement dynamics, we use the FOTOC given by $C(t) \equiv 1 - \real{\langle \hat{W}^\dagger_\delta(t) \hat{V}^\dagger(0) \hat{W}_\delta(t) \hat{V}(0)\rangle}$, where $\hat{W}_\delta$ is an arbitrary rotation with generator $\hat{S}_\alpha$, and $\hat{V} = \ket{\psi_0}\bra{\psi_0}$ is the projector on the initial state $\ket{\psi_0}$. In this paper, we use Bloch coherent states as the initial states, and choose the generator whose expectation value $\avg{\hat{S}_\alpha}$ is maximised under the initial state.

We choose to work with Bloch coherent states since, as symmetric minimum-uncertainty states, they provide a somewhat localised probe of phase space.  They are also relatively easy to generate experimentally, corresponding to a rotation of the lowest-eigenvalue eigenstate of $\hat{J}_z$ to arbitrary coordinates $(z,\phi)$ on the Bloch sphere~\cite{Arecchi1972}.

We work with sufficiently small $\delta \ll 1$ such that the FOTOC simplifies to the variance of the generator $\hat{S}_\alpha$, as $C(t) \approx \delta^2 \var{[\hat{S}_\alpha(t)]} + O(\delta^3)$~\cite{Schmitt2019, Lewis-Swan2019}. The predicted value of the FOTOC for the infinite temperature UDE, $\hat{\rho} = \mathbb{I}/(N+1)$ %, where $\mathbb{I}$ the identity,
in the Bose-Hubbard dimer is~\cite{Kidd2020}
\begin{eqnarray}
    C_{\rm UDE} &=& \delta^2\left[ \Tr{(\hat{\rho} \hat{S}_\alpha^2)} - \Tr{(\hat{\rho} \hat{S}_\alpha)}^2 \right] \nonumber \\
    &=& \delta^2\left[ \Tr{(\hat{S}_\alpha^2)}/(N+1) - \Tr{(\hat{S}_\alpha)}^2/(N+1)^2 \right] \nonumber \\
    &=& \delta^2N(N+2)/12.
\end{eqnarray}

\subsection{Driven dimer FOTOC dynamics}

Figure~\ref{fig:FOTOC}(a) shows the dynamics of the FOTOC in the saddle-point and chaotic regimes. In contrast to the behaviour near a stable fixed point, where the FOTOC remains small and bounded, here the FOTOCs grow exponentially until they approach the respective values predicted by the diagonal ensemble.  Semiclassical arguments indicate that the growth rate $\lambda_Q$ of this FOTOC should be related to the  classical Lyapunov exponent as $\lambda_Q = 4\lambda$~\footnote{That the FOTOC corresponds to a variance, $\var{(\hat{S}_x)}$ gives rise to one factor of two~\cite{Lewis-Swan2019}.  Additionally, the initial phase-space trajectories are orthogonal to the $ \langle \hat{S}_x\rangle$-axis.  The change in $ \langle \hat{S}_x\rangle$ is thus a second-order effect, which gives rise to a second factor of two.}
Here, we numerically determine the saddle point FOTOC growth rate to be $\lambda_Q = 4 \times 1.90(5)$, consistent with the saddle-point exponent $\lambda = 2$, determined via linear stability analysis.  The exponential growth is arrested at an Ehrenfest time $t_E$.  For the saddle-driven scrambling, $t_E \approx 2.0$, which is consistent with the prediction given by $t_E \sim (2\lambda)^{-1} \ln{(N)} \approx 1.73$.
\label{DDFOTOC}

For the driven, chaotic case, the pre-Ehrenfest dynamics is a little more complicated, with the exponential growth preceded by a short time of slow growth (seen clearly in, e.g., Fig. 2(d) of Ref.~\cite{Kidd2020}).
Furthermore, the numerically determined growth rate during the exponential phase $\lambda_Q = 4 \times 0.92(2)$ is approximately four times higher than what would be expected from the corresponding classical Lyapunov exponent,
$\lambda \approx 0.2094(4)$, determined numerically via the tangent-space method~\cite{Chavez-Carlos2016, Skokos2010, Lewis-Swan2019, Zeni1995}.

Despite both the saddle-point and the chaotic regimes showing scrambling behaviour in the short-time dynamics, there is a marked difference in the long-time behaviours.  The saddle point FOTOCs exhibit considerable variability around the diagonal-ensemble predictions.  This variability is especially marked for an initial state centred directly on the saddle point, where it is manifest in large oscillations. These oscillations occurred for all system sizes we simulated ($N=10^1$, $N=10^2$, $N=10^3$) and are comparable to the maximum values of the FOTOC. As we discuss further below, these oscillations are associated with (near) revivals of the initial state. In contrast, in the chaotic regime the FOTOC does not exhibit large oscillations but saturates to the uniform diagonal ensemble~\footnote{We note some regularly-spaced `pulses' in the OTOC over intermediate time-scales.  The timing of this transient feature seems to correlate with the driving period.} prediction with only small-amplitude fluctuations for sufficiently large $N$ as predicted by the eigenstate thermalisation hypothesis~\cite{Rigol2008}.

\subsection{Entanglement Entropy}

The differences between the chaotic and the saddle point regimes are even more marked in the dynamics of the entanglement entropy.  Here we calculate the von Neumann entropy of the reduced density matrix $\hat{\rho}_{s}$ obtained after tracing out $N-s$ particles~\cite{Frowis2015}:
$\mathcal{S}_\text{E} (\hat{\rho}_{s}) = -\Tr{[\hat{\rho}_s \ln{(\hat{\rho}_{s})}]}$.
Figure~\ref{fig:FOTOC}(b) shows the half-system ($s=N/2$) entanglement entropy as a function of time.
In the chaotic regime, the entanglement entropy grows almost monotonically, and shows very little variation once it saturates at the Page curve prediction.  In contrast, the saddle point entanglement entropy exhibits markedly periodic oscillations with distinct revivals. The size of the oscillations means that the system periodically almost completely disentangles. These disentangling points coincide with time at which $C(t)\approx 0$, illustrating the close connection between the dynamics of entanglement entropy and the FOTOCs previously observed in Ref.~\cite{Lewis-Swan2019}.

The lack of fluctuations in the entanglement entropy after the Ehrenfest time means that its long-time value is very well defined, which allows us to test its scaling with system size.  To obtain the results in Fig.~\ref{fig:FOTOC}(c), we time-average the entropy after saturation, over the period $20 \leq J_0 t \leq 200$. The vertical lines indicate the uncertainty.
For small subsystems, the chaotic post-Ehrenfest entanglement entropy matches that of the infinite-temperature uniform diagonal ensemble, but diverges for larger subsystems. The volume law scaling of the chaotic state entanglement entropy matches the Page curve, the average entanglement entropy of a subsystem given the whole system is in a random pure state~\cite{Bianchi2019}. For a bipartite system with subsystems $A,B$ of Hilbert space dimensions $d_{A,B}$ satisfying $d_A \leq d_B$, this quantity is
\begin{equation}
    \avg{\mathcal{S}_\text{E}(\hat{\rho}_A)} = \Psi{(d_A d_B + 1)} - \Psi{(d_B + 1)} - \frac{d_A - 1}{2d_B},
\end{equation}
where $\Psi(x) = \Gamma'(x)/\Gamma(x)$ is the digamma function.  In the dimer case, $d_A = s+1$ and $d_B = N-s+1$, where $s \in [0, N/2]$ is the number of particles in subsystem $A$.

\subsection{Diagonal distribution and thermalisation}

The qualitative differences in the long-time dynamics of the two regimes can be traced back to the eigenstructure, which we probe through the diagonal entropy~\cite{DAlessio2016}, defined as the von Neumann entropy of the diagonal ensemble,
\begin{equation} \label{eq:Shannon1}
    \mathcal{S}_\text{D}(\psi) = \sum_n \abs{\braket{\psi}{\Phi_n}}^2 \ln{\left(\abs{\braket{\psi}{\Phi_n}}^2 \right)},
\end{equation}
with initial coherent state $\ket{\psi}$ and eigenstates $\ket{\Phi_n}$. The diagonal entropy quantifies the delocalisation of the state $\ket{\psi}$ in the basis of energy eigenstates, or Floquet modes in the case of a Floquet system. The Gaussian orthogonal ensemble (GOE) of random matrices gives a prediction of $\mathcal{S}_\text{GOE} \approx \ln{[0.48(N+1)]}$~\cite{DAlessio2014} for fully thermalising states under the Bose-Hubbard Hamiltonian~\eqref{eq:H}, which is real and symmetric.

The diagonal distributions of the states used in the chaotic and saddle-point regimes are shown in Fig.~\ref{fig:diag_ensembles}(a), with the corresponding diagonal entropies in Fig.~\ref{fig:diag_ensembles}(c). The chaotic state matches the GOE prediction, indicating that a large number of eigenstates participate in the FOTOC dynamics and that the state will thermalise. In contrast, the diagonal entropy of the saddle point state is much smaller than the GOE prediction, indicating that relatively few eigenstates participate in the FOTOC dynamics and that the state will not thermalise. The three dominant eigenstates are illustrated in Fig.~\ref{fig:diag_ensembles}(d) and are all highly localised around the saddle point. The saddle-perturbed state has intermediate diagonal entropy and thus exhibits small amplitude FOTOC oscillations without distinct revivals or saturation. The diagonal distribution of a state centred on a stable fixed point is almost a delta distribution ($\mathcal{S}_\text{D} \approx 0$) and therefore the state's FOTOC and entanglement entropy are predicted to be nearly time-independent, which matches our observations.

The connection among the diagonal ensemble and the FOTOC and entanglement dynamics is clearly revealed in the Fourier spectra of these quantities, which effectively functions as a probe of chaotic behaviour~\cite{Fortes2019}. As evident in Fig.~\ref{fig:FOTOC_fft}(a), the saddle-point FOTOC spectrum displays distinct peaks at frequencies corresponding to the transition energies between eigenstates that dominate the diagonal ensemble. These eigenstates can be identified by their distinctly high overlaps with the evolving state, shown in Fig.~\ref{fig:diag_ensembles}(b).

The Fourier spectrum of the entanglement entropy displays many of the same features as that of the FOTOC for $N = 100$. The Fourier spectrum of the saddle-point entropy has few, sharp peaks, whereas those of the chaotic phase space and saddle-perturbed state exhibit no distinct peaks. The appearance of beating in the saddle-point entanglement entropy dynamics is evident in the double-peaked Fourier spectrum, giving oscillation period $T_o \approx 4.00/J_0$ and beat period $T_b \approx 106/J_0$. The beat period can be interpreted as the partial revival timescale. Periodic partial revivals are similarly evident in the long-time dynamics of the $N=1000$ saddle point FOTOC in Fig.~\ref{fig:FOTOC}(a).

That so few eigenstates participate in the saddle-dominated dynamics is the reason why these regular revivals are so substantial. The revival dynamics is illustrated qualitatively in the phase-space $Q$ distribution, shown in Fig.~\ref{fig:FOTOC_fft}(c). The coherent state, initially located on the saddle point, is periodically sheared along the separatrix and extends around the Bloch sphere before the separate $Q$ distribution arms recombine at the saddle point, leading to a fairly complete reconstruction of the initial coherent state. This periodic dividing and recombining is the source of the oscillatory FOTOC behaviour.

\subsection{Semiclassical simulations for large \textit{N}}

\begin{figure}[ht]
    \centering
    \includegraphics[width=\columnwidth]{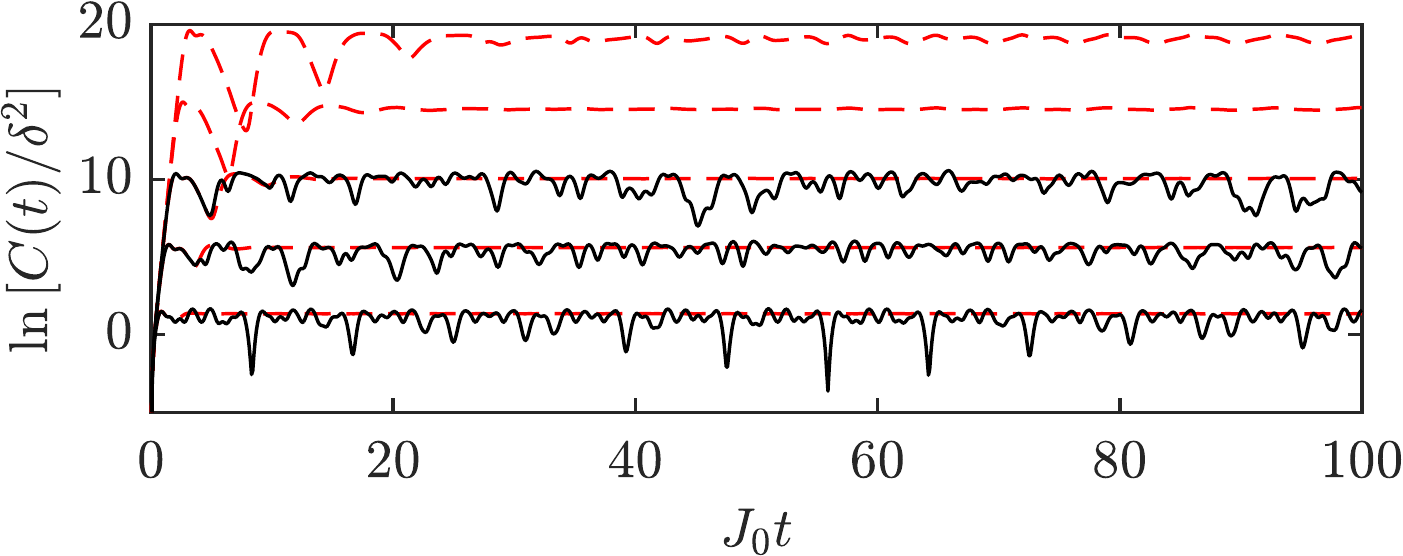}
    \caption{\label{fig:FOTOC_TW}Fidelity out-of-time-order correlators (FOTOCs) for the saddle point in Fig.~\ref{fig:FOTOC} for $N=10^p$ particles, where $p=1--5$ from the bottom. Exact dynamics are indicated by black solid lines and truncated Wigner approximations are indicated by red dashed lines. Exponential short-time FOTOC growth is evident and the Ehrenfest (scrambling) time increases with $N$.}
\end{figure}

We have thus far solved the driven-dimer dynamics exactly, using exact diagonalisation.  For larger system sizes, we make use of the truncated Wigner method, an approximate approach that uses an ensemble of stochastically sampled semiclassical trajectories~\cite{Walls2008, Olsen2009, Polkovnikov2010, Ruostekoski2013}. Most accurate when the number of particles is large, the method has been successfully applied in a range of quantum optics and ultracold atoms systems, including FOTOC dynamics in LMG and Dicke models~\cite{Pilatowsky-Cameo2020}.

The Wigner simulations can be benchmarked against the exact simulations for small $N$.
As shown in Fig.~\ref{fig:FOTOC_TW}, the Wigner method correctly predicts the rapid exponential growth of the FOTOC, and some transient oscillations immediately after the Ehrenfest time. However, it fails to reproduce the persistent fluctuations seen in the long-time dynamics.  Thus, reliance on the Wigner method alone could give a false prediction of thermalisation after the Ehrenfest time.
The failure of the Wigner method is consistent with the idea that these persistent oscillations are due to beating between a few eigenstates: superpositions of macroscopically distinct states cannot be accurately sampled with the truncated Wigner method due to the nonpositive fringes in a Wigner function that arise from interference.

For short times, when the truncated Wigner method is demonstrably reliable, we see from Fig.~\ref{fig:FOTOC_TW} that the OTOC continues to grow exponentially until it reaches the diagonal-ensemble prediction, which increases with $N$.  The corresponding Ehrenfest time, thus, also increases with $N$, consistent with the logarithmic dependence given in Sec.~\ref{DDFOTOC}.

\subsection{Dicke FOTOC dynamics}

\begin{figure}[ht]
    \centering
    \includegraphics[width=\columnwidth]{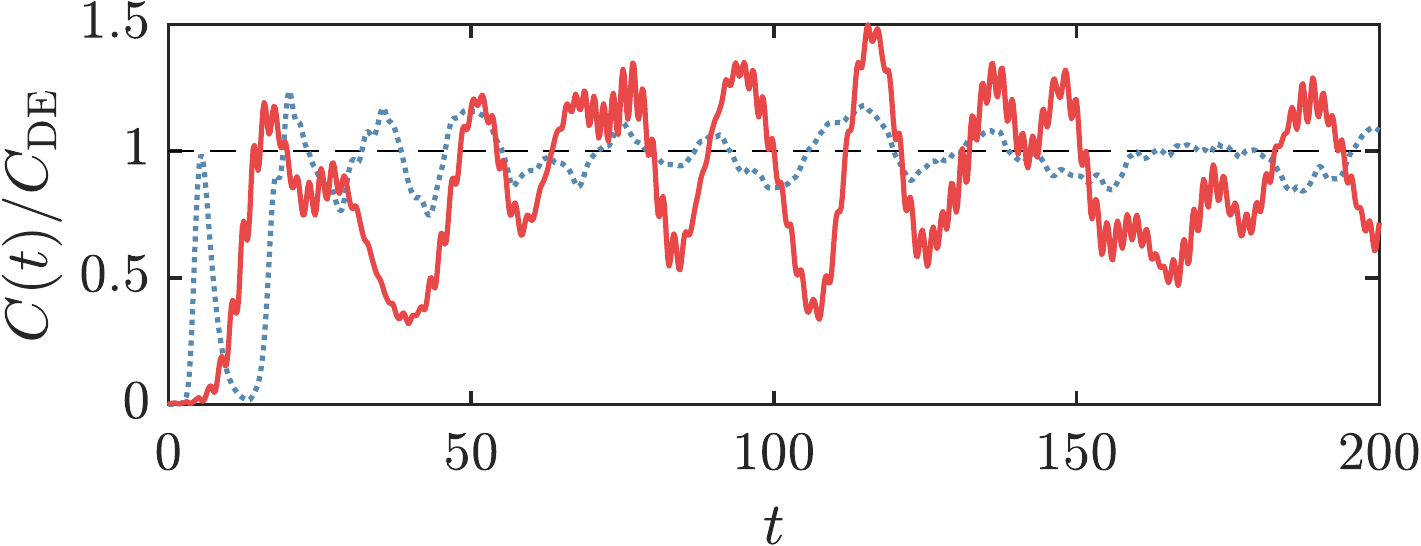}
    \caption{\label{fig:dicke_FOTOC}a) FOTOCs for the Dicke-model saddle point for strong (blue dotted line)  and weak (red solid line) atom-field couplings, scaled by their respective diagonal ensemble predictions; $C_\text{DE}/C_\text{UDE} \approx 0.54$ and $C_\text{DE}/C_\text{UDE} \approx 0.027$ for the strongly and weakly coupled regimes, respectively. The strongly coupled regime corresponds to $\Delta = 0.5$ ($\gamma/\gamma_c \approx 2.3$) and the weakly coupled regime corresponds to $\Delta = 3$ ($\gamma/\gamma_c \approx 1.1$). In both regimes $N=200$, $\gamma = 0.66$ and $\omega = 0.5$.}
\end{figure}

Finally, we use the Dicke model to explore whether the characteristic differences between saddle-driven and chaos-driven scrambling described above survive in a system with higher-dimensional phase space, and without the need for driving to induce chaos.  We consider the Dicke model in two parameter regimes: a weakly coupled regime with $\gamma/\gamma_c \approx 1.1$ and a strongly coupled regime with $\gamma/\gamma_c \approx 2.3$.

The weakly coupled regime is just above the critical value of atom field coupling and the phase-space energy shell of the saddle point features no discernible chaos. The strongly coupled regime is well above the critical value of atom field coupling and features widespread chaos in the saddle-point energy shell.
As in the Dicke model, chaos in the semiclassical system is reflected in delocalised eigenstates in the quantum system:
The diagonal entropy of a coherent state localised at the saddle point in the Dicke model is $\mathcal{S}_\text{D} \approx 4.0$ for the strongly coupled regime and $\mathcal{S}_\text{D} \approx 1.7$ for the weakly coupled regime~\footnote{We quote unnormalised values since there is no upper bound to the diagonal entropy in the infinite-dimensional Hilbert space of the Dicke model.}.

The dynamics of FOTOCs for the Dicke model saddle point in the strongly and weakly coupled regimes are shown in Fig.~\ref{fig:dicke_FOTOC}.
The FOTOCs for both regimes grow exponentially before fluctuating about their respective diagonal ensemble prediction.  Just as in the driven dimer case, there is a substantial quantitative difference in the diagonal ensemble predictions between the chaos-dominated and the saddle-dominated regime: $C_\text{DE}/C_\text{UDE} \approx 0.54$ in the case of strong coupling and $C_\text{DE}/C_\text{UDE} \approx 0.027$ for weak coupling, where for convenience we have normalised by the value for the uniform diagonal ensemble.
We note in passing that the diagonal ensemble results are similar to the respective microcanonical predictions for the FOTOCs, particularly in the strongly coupled case. The microcanonical ensemble FOTOCs, calculated by averaging FOTOC predictions over eigenstates within one standard deviation of the saddle point state energy, are $C_\text{ME}/C_\text{UDE} \approx 0.55$ (252 eigenstates) and $C_\text{ME}/C_\text{UDE} \approx 0.021$ (38 eigenstates) for the strongly and weakly coupled regimes, respectively.

In contrast to the driven dimer FOTOC, the qualitative features of the long-time dynamics are not so clear-cut in the Dicke model: The persistent oscillations in the saddle-dominated FOTOC are not as regular as in the dimer case, and the chaos-dominated FOTOC has persistent fluctuations past the Ehrenfest time.  The latter suggests that the chaotic system does not thermalise completely over the relatively long time scales we simulated.

Nevertheless, it is the case that, in the chaos-dominated, strongly coupled regime, the FOTOC exhibits relatively smaller oscillations than in the weakly coupled regime, as predicted by its larger diagonal entropy.

\section{Conclusions}

We studied the long-time dynamics of the fidelity out-of-time-order correlator and entanglement entropy for simple systems of ultracold atoms in the presence of chaos and in the vicinity of a saddle point.  The main conclusion to be drawn is that, although saddle points and semiclassical chaos both drive exponential growth of OTOCs, these two causes of scrambling can lead to very different long-time behaviour.  In particular, we find that, without chaos, a state located on or very near to the saddle point fails to thermalise. The Bose-Hubbard dimer saddle-point FOTOC and entanglement entropy dynamics were distinctly periodic, with dominant frequencies given by transition energies between eigenstates with high overlap on the evolving state. In the Dicke model and in the driven dimer displaced slightly from the saddle point, the FOTOC and entanglement entropy dynamics were aperiodic but, nevertheless, failed to thermalise completely, exhibiting persistent oscillations.  In any case, the diagonal entropy of the initial state correctly predicted the degree of thermalisation regardless of short-time FOTOC behaviour, suggesting that it is the degree of localisation in the energy eigenbasis that determines the long-time behaviour.

In order to distinguish clearly the different characteristics of the chaos-induced thermalisation versus saddle-point-driven scrambling, we have chosen regimes for the Bose-Hubbard dimer where there is either a saddle point or chaos in the semiclassical phase space, but not both.  For regimes where both are present, such as in the Dicke model, the OTOC growth may reflect an interplay of the two such that the long-term thermal behaviour arises from the chaotic nature of the system, but where the initial rapid growth in correlations is determined by the saddle point.

\bibliographystyle{apsrev4-2}
% \bibliography{references.bib}

\begin{thebibliography}{53}%
\makeatletter
\providecommand \@ifxundefined [1]{%
 \@ifx{#1\undefined}
}%
\providecommand \@ifnum [1]{%
 \ifnum #1\expandafter \@firstoftwo
 \else \expandafter \@secondoftwo
 \fi
}%
\providecommand \@ifx [1]{%
 \ifx #1\expandafter \@firstoftwo
 \else \expandafter \@secondoftwo
 \fi
}%
\providecommand \natexlab [1]{#1}%
\providecommand \enquote  [1]{``#1''}%
\providecommand \bibnamefont  [1]{#1}%
\providecommand \bibfnamefont [1]{#1}%
\providecommand \citenamefont [1]{#1}%
\providecommand \href@noop [0]{\@secondoftwo}%
\providecommand \href [0]{\begingroup \@sanitize@url \@href}%
\providecommand \@href[1]{\@@startlink{#1}\@@href}%
\providecommand \@@href[1]{\endgroup#1\@@endlink}%
\providecommand \@sanitize@url [0]{\catcode `\\12\catcode `\$12\catcode
  `\&12\catcode `\#12\catcode `\^12\catcode `\_12\catcode `\%12\relax}%
\providecommand \@@startlink[1]{}%
\providecommand \@@endlink[0]{}%
\providecommand \url  [0]{\begingroup\@sanitize@url \@url }%
\providecommand \@url [1]{\endgroup\@href {#1}{\urlprefix }}%
\providecommand \urlprefix  [0]{URL }%
\providecommand \Eprint [0]{\href }%
\providecommand \doibase [0]{https://doi.org/}%
\providecommand \selectlanguage [0]{\@gobble}%
\providecommand \bibinfo  [0]{\@secondoftwo}%
\providecommand \bibfield  [0]{\@secondoftwo}%
\providecommand \translation [1]{[#1]}%
\providecommand \BibitemOpen [0]{}%
\providecommand \bibitemStop [0]{}%
\providecommand \bibitemNoStop [0]{.\EOS\space}%
\providecommand \EOS [0]{\spacefactor3000\relax}%
\providecommand \BibitemShut  [1]{\csname bibitem#1\endcsname}%
\let\auto@bib@innerbib\@empty
%</preamble>
\bibitem [{\citenamefont {Larkin}\ and\ \citenamefont
  {Ovchinnikov}(1968)}]{Larkin1969}%
  \BibitemOpen
  \bibfield  {author} {\bibinfo {author} {\bibfnamefont {A.~I.}\ \bibnamefont
  {Larkin}}\ and\ \bibinfo {author} {\bibfnamefont {Y.~N.}\ \bibnamefont
  {Ovchinnikov}},\ }\href@noop {} {\bibfield  {journal} {\bibinfo  {journal}
  {Zh. Eksp. Teor. Fiz.}\ }\textbf {\bibinfo {volume} {55}},\ \bibinfo {pages}
  {2262} (\bibinfo {year} {1968})}\BibitemShut {NoStop}%
\bibitem [{\citenamefont {Larkin}\ and\ \citenamefont
  {Ovchinnikov}(1969)}]{Larkin1969_2}%
  \BibitemOpen
  \bibfield  {author} {\bibinfo {author} {\bibfnamefont {A.~I.}\ \bibnamefont
  {Larkin}}\ and\ \bibinfo {author} {\bibfnamefont {Y.~N.}\ \bibnamefont
  {Ovchinnikov}},\ }\href {http://www.jetp.ac.ru/cgi-bin/dn/e_028_06_1200.pdf}
  {\bibfield  {journal} {\bibinfo  {journal} {J. Exp. Theor. Phys.}\ }\textbf
  {\bibinfo {volume} {28}},\ \bibinfo {pages} {1200} (\bibinfo {year}
  {1969})}\BibitemShut {NoStop}%
\bibitem [{\citenamefont {Hayden}\ and\ \citenamefont
  {Preskill}(2007)}]{Hayden2007}%
  \BibitemOpen
  \bibfield  {author} {\bibinfo {author} {\bibfnamefont {P.}~\bibnamefont
  {Hayden}}\ and\ \bibinfo {author} {\bibfnamefont {J.}~\bibnamefont
  {Preskill}},\ }\href {https://doi.org/10.1088/1126-6708/2007/09/120}
  {\bibfield  {journal} {\bibinfo  {journal} {J. High Energy Phys.}\ }\textbf
  {\bibinfo {volume} {2007}}\bibinfo  {number} { (09)},\ \bibinfo {pages}
  {120}}\BibitemShut {NoStop}%
\bibitem [{\citenamefont {Sekino}\ and\ \citenamefont
  {Susskind}(2008)}]{Sekino2008}%
  \BibitemOpen
\bibfield  {number} {  }\bibfield  {author} {\bibinfo {author} {\bibfnamefont
  {Y.}~\bibnamefont {Sekino}}\ and\ \bibinfo {author} {\bibfnamefont
  {L.}~\bibnamefont {Susskind}},\ }\href
  {https://doi.org/10.1088/1126-6708/2008/10/065} {\bibfield  {journal}
  {\bibinfo  {journal} {J. High Energy Phys.}\ }\textbf {\bibinfo {volume}
  {2008}}\bibinfo  {number} { (10)},\ \bibinfo {pages} {065}}\BibitemShut
  {NoStop}%
\bibitem [{\citenamefont {Shenker}\ and\ \citenamefont
  {Stanford}(2014)}]{Shenker2014}%
  \BibitemOpen
\bibfield  {number} {  }\bibfield  {author} {\bibinfo {author} {\bibfnamefont
  {S.~H.}\ \bibnamefont {Shenker}}\ and\ \bibinfo {author} {\bibfnamefont
  {D.}~\bibnamefont {Stanford}},\ }\href
  {https://doi.org/10.1007/JHEP12(2014)046} {\bibfield  {journal} {\bibinfo
  {journal} {J. High Energy Phys.}\ }\textbf {\bibinfo {volume} {2014}}\bibinfo
   {number} { (12)},\ \bibinfo {pages} {46}}\BibitemShut {NoStop}%
\bibitem [{\citenamefont {Roberts}\ \emph {et~al.}(2015)\citenamefont
  {Roberts}, \citenamefont {Stanford},\ and\ \citenamefont
  {Susskind}}]{Roberts2015}%
  \BibitemOpen
\bibfield  {number} {  }\bibfield  {author} {\bibinfo {author} {\bibfnamefont
  {D.~A.}\ \bibnamefont {Roberts}}, \bibinfo {author} {\bibfnamefont
  {D.}~\bibnamefont {Stanford}},\ and\ \bibinfo {author} {\bibfnamefont
  {L.}~\bibnamefont {Susskind}},\ }\href
  {https://doi.org/10.1007/JHEP03(2015)051} {\bibfield  {journal} {\bibinfo
  {journal} {J. High Energy Phys.}\ }\textbf {\bibinfo {volume} {2015}}\bibinfo
   {number} { (3)},\ \bibinfo {pages} {51}}\BibitemShut {NoStop}%
\bibitem [{\citenamefont {Hosur}\ \emph {et~al.}(2016)\citenamefont {Hosur},
  \citenamefont {Qi}, \citenamefont {Roberts},\ and\ \citenamefont
  {Yoshida}}]{Hosur2016}%
  \BibitemOpen
\bibfield  {number} {  }\bibfield  {author} {\bibinfo {author} {\bibfnamefont
  {P.}~\bibnamefont {Hosur}}, \bibinfo {author} {\bibfnamefont {X.-L.}\
  \bibnamefont {Qi}}, \bibinfo {author} {\bibfnamefont {D.~A.}\ \bibnamefont
  {Roberts}},\ and\ \bibinfo {author} {\bibfnamefont {B.}~\bibnamefont
  {Yoshida}},\ }\href {https://doi.org/10.1007/JHEP02(2016)004} {\bibfield
  {journal} {\bibinfo  {journal} {J. High Energy Phys.}\ }\textbf {\bibinfo
  {volume} {2016}}\bibinfo  {number} { (2)},\ \bibinfo {pages} {4}}\BibitemShut
  {NoStop}%
\bibitem [{\citenamefont {Maldacena}\ \emph {et~al.}(2016)\citenamefont
  {Maldacena}, \citenamefont {Shenker},\ and\ \citenamefont
  {Stanford}}]{Maldacena2016}%
  \BibitemOpen
\bibfield  {number} {  }\bibfield  {author} {\bibinfo {author} {\bibfnamefont
  {J.}~\bibnamefont {Maldacena}}, \bibinfo {author} {\bibfnamefont {S.~H.}\
  \bibnamefont {Shenker}},\ and\ \bibinfo {author} {\bibfnamefont
  {D.}~\bibnamefont {Stanford}},\ }\href
  {https://doi.org/10.1007/JHEP08(2016)106} {\bibfield  {journal} {\bibinfo
  {journal} {J. High Energy Phys.}\ }\textbf {\bibinfo {volume} {2016}}\bibinfo
   {number} { (8)},\ \bibinfo {pages} {106}}\BibitemShut {NoStop}%
\bibitem [{\citenamefont {Rozenbaum}\ \emph {et~al.}(2017)\citenamefont
  {Rozenbaum}, \citenamefont {Ganeshan},\ and\ \citenamefont
  {Galitski}}]{Rozenbaum2017}%
  \BibitemOpen
\bibfield  {number} {  }\bibfield  {author} {\bibinfo {author} {\bibfnamefont
  {E.~B.}\ \bibnamefont {Rozenbaum}}, \bibinfo {author} {\bibfnamefont
  {S.}~\bibnamefont {Ganeshan}},\ and\ \bibinfo {author} {\bibfnamefont
  {V.}~\bibnamefont {Galitski}},\ }\href
  {https://doi.org/10.1103/PhysRevLett.118.086801} {\bibfield  {journal}
  {\bibinfo  {journal} {Phys. Rev. Lett.}\ }\textbf {\bibinfo {volume} {118}},\
  \bibinfo {pages} {086801} (\bibinfo {year} {2017})}\BibitemShut {NoStop}%
\bibitem [{\citenamefont {Rammensee}\ \emph {et~al.}(2018)\citenamefont
  {Rammensee}, \citenamefont {Urbina},\ and\ \citenamefont
  {Richter}}]{Rammensee2018}%
  \BibitemOpen
  \bibfield  {author} {\bibinfo {author} {\bibfnamefont {J.}~\bibnamefont
  {Rammensee}}, \bibinfo {author} {\bibfnamefont {J.~D.}\ \bibnamefont
  {Urbina}},\ and\ \bibinfo {author} {\bibfnamefont {K.}~\bibnamefont
  {Richter}},\ }\href {https://doi.org/10.1103/PhysRevLett.121.124101}
  {\bibfield  {journal} {\bibinfo  {journal} {Phys. Rev. Lett.}\ }\textbf
  {\bibinfo {volume} {121}},\ \bibinfo {pages} {124101} (\bibinfo {year}
  {2018})}\BibitemShut {NoStop}%
\bibitem [{\citenamefont {Garc{\'{i}}a-Mata}\ \emph {et~al.}(2018)\citenamefont
  {Garc{\'{i}}a-Mata}, \citenamefont {Saraceno}, \citenamefont {Jalabert},
  \citenamefont {Roncaglia},\ and\ \citenamefont
  {Wisniacki}}]{Garcia-Mata2018}%
  \BibitemOpen
  \bibfield  {author} {\bibinfo {author} {\bibfnamefont {I.}~\bibnamefont
  {Garc{\'{i}}a-Mata}}, \bibinfo {author} {\bibfnamefont {M.}~\bibnamefont
  {Saraceno}}, \bibinfo {author} {\bibfnamefont {R.~A.}\ \bibnamefont
  {Jalabert}}, \bibinfo {author} {\bibfnamefont {A.~J.}\ \bibnamefont
  {Roncaglia}},\ and\ \bibinfo {author} {\bibfnamefont {D.~A.}\ \bibnamefont
  {Wisniacki}},\ }\href {https://doi.org/10.1103/PhysRevLett.121.210601}
  {\bibfield  {journal} {\bibinfo  {journal} {Phys. Rev. Lett.}\ }\textbf
  {\bibinfo {volume} {121}},\ \bibinfo {pages} {210601} (\bibinfo {year}
  {2018})}\BibitemShut {NoStop}%
\bibitem [{\citenamefont {Lewis-Swan}\ \emph {et~al.}(2019)\citenamefont
  {Lewis-Swan}, \citenamefont {Safavi-Naini}, \citenamefont {Kaufman},\ and\
  \citenamefont {Rey}}]{Lewis-Swan2019}%
  \BibitemOpen
  \bibfield  {author} {\bibinfo {author} {\bibfnamefont {R.~J.}\ \bibnamefont
  {Lewis-Swan}}, \bibinfo {author} {\bibfnamefont {A.}~\bibnamefont
  {Safavi-Naini}}, \bibinfo {author} {\bibfnamefont {A.~M.}\ \bibnamefont
  {Kaufman}},\ and\ \bibinfo {author} {\bibfnamefont {A.~M.}\ \bibnamefont
  {Rey}},\ }\href {https://doi.org/10.1038/s42254-019-0090-y} {\bibfield
  {journal} {\bibinfo  {journal} {Nat. Rev. Phys.}\ }\textbf {\bibinfo {volume}
  {1}},\ \bibinfo {pages} {627} (\bibinfo {year} {2019})}\BibitemShut {NoStop}%
\bibitem [{\citenamefont {Prakash}\ and\ \citenamefont
  {Lakshminarayan}(2020)}]{Prakash2020}%
  \BibitemOpen
  \bibfield  {author} {\bibinfo {author} {\bibfnamefont {R.}~\bibnamefont
  {Prakash}}\ and\ \bibinfo {author} {\bibfnamefont {A.}~\bibnamefont
  {Lakshminarayan}},\ }\href {https://doi.org/10.1103/PhysRevB.101.121108}
  {\bibfield  {journal} {\bibinfo  {journal} {Phys. Rev. B}\ }\textbf {\bibinfo
  {volume} {101}},\ \bibinfo {pages} {121108(R)} (\bibinfo {year}
  {2020})}\BibitemShut {NoStop}%
\bibitem [{\citenamefont {Xu}\ \emph {et~al.}(2020)\citenamefont {Xu},
  \citenamefont {Scaffidi},\ and\ \citenamefont {Cao}}]{Xu2020}%
  \BibitemOpen
  \bibfield  {author} {\bibinfo {author} {\bibfnamefont {T.}~\bibnamefont
  {Xu}}, \bibinfo {author} {\bibfnamefont {T.}~\bibnamefont {Scaffidi}},\ and\
  \bibinfo {author} {\bibfnamefont {X.}~\bibnamefont {Cao}},\ }\href
  {https://doi.org/10.1103/PhysRevLett.124.140602} {\bibfield  {journal}
  {\bibinfo  {journal} {Phys. Rev. Lett.}\ }\textbf {\bibinfo {volume} {124}},\
  \bibinfo {pages} {140602} (\bibinfo {year} {2020})}\BibitemShut {NoStop}%
\bibitem [{\citenamefont {G{\"{a}}rttner}\ \emph {et~al.}(2017)\citenamefont
  {G{\"{a}}rttner}, \citenamefont {Bohnet}, \citenamefont {Safavi-Naini},
  \citenamefont {Wall}, \citenamefont {Bollinger},\ and\ \citenamefont
  {Rey}}]{Garttner2017}%
  \BibitemOpen
  \bibfield  {author} {\bibinfo {author} {\bibfnamefont {M.}~\bibnamefont
  {G{\"{a}}rttner}}, \bibinfo {author} {\bibfnamefont {J.~G.}\ \bibnamefont
  {Bohnet}}, \bibinfo {author} {\bibfnamefont {A.}~\bibnamefont
  {Safavi-Naini}}, \bibinfo {author} {\bibfnamefont {M.~L.}\ \bibnamefont
  {Wall}}, \bibinfo {author} {\bibfnamefont {J.~J.}\ \bibnamefont
  {Bollinger}},\ and\ \bibinfo {author} {\bibfnamefont {A.~M.}\ \bibnamefont
  {Rey}},\ }\href {https://doi.org/10.1038/nphys4119} {\bibfield  {journal}
  {\bibinfo  {journal} {Nat. Phys.}\ }\textbf {\bibinfo {volume} {13}},\
  \bibinfo {pages} {781} (\bibinfo {year} {2017})}\BibitemShut {NoStop}%
\bibitem [{\citenamefont {G{\"{a}}rttner}\ \emph {et~al.}(2018)\citenamefont
  {G{\"{a}}rttner}, \citenamefont {Hauke},\ and\ \citenamefont
  {Rey}}]{Garttner2018}%
  \BibitemOpen
  \bibfield  {author} {\bibinfo {author} {\bibfnamefont {M.}~\bibnamefont
  {G{\"{a}}rttner}}, \bibinfo {author} {\bibfnamefont {P.}~\bibnamefont
  {Hauke}},\ and\ \bibinfo {author} {\bibfnamefont {A.~M.}\ \bibnamefont
  {Rey}},\ }\href {https://doi.org/10.1103/PhysRevLett.120.040402} {\bibfield
  {journal} {\bibinfo  {journal} {Phys. Rev. Lett.}\ }\textbf {\bibinfo
  {volume} {120}},\ \bibinfo {pages} {040402} (\bibinfo {year}
  {2018})}\BibitemShut {NoStop}%
\bibitem [{\citenamefont {Pilatowsky-Cameo}\ \emph {et~al.}(2020)\citenamefont
  {Pilatowsky-Cameo}, \citenamefont {Ch{\'{a}}vez-Carlos}, \citenamefont
  {Bastarrachea-Magnani}, \citenamefont {Str{\'{a}}nsk{\'{y}}}, \citenamefont
  {Lerma-Hern{\'{a}}ndez}, \citenamefont {Santos},\ and\ \citenamefont
  {Hirsch}}]{Pilatowsky-Cameo2020}%
  \BibitemOpen
  \bibfield  {author} {\bibinfo {author} {\bibfnamefont {S.}~\bibnamefont
  {Pilatowsky-Cameo}}, \bibinfo {author} {\bibfnamefont {J.}~\bibnamefont
  {Ch{\'{a}}vez-Carlos}}, \bibinfo {author} {\bibfnamefont {M.~A.}\
  \bibnamefont {Bastarrachea-Magnani}}, \bibinfo {author} {\bibfnamefont
  {P.}~\bibnamefont {Str{\'{a}}nsk{\'{y}}}}, \bibinfo {author} {\bibfnamefont
  {S.}~\bibnamefont {Lerma-Hern{\'{a}}ndez}}, \bibinfo {author} {\bibfnamefont
  {L.~F.}\ \bibnamefont {Santos}},\ and\ \bibinfo {author} {\bibfnamefont
  {J.~G.}\ \bibnamefont {Hirsch}},\ }\href
  {https://doi.org/10.1103/PhysRevE.101.010202} {\bibfield  {journal} {\bibinfo
   {journal} {Phys. Rev. E}\ }\textbf {\bibinfo {volume} {101}},\ \bibinfo
  {pages} {010202(R)} (\bibinfo {year} {2020})}\BibitemShut {NoStop}%
\bibitem [{\citenamefont {Kidd}\ \emph {et~al.}(2020)\citenamefont {Kidd},
  \citenamefont {Safavi-Naini},\ and\ \citenamefont {Corney}}]{Kidd2020}%
  \BibitemOpen
  \bibfield  {author} {\bibinfo {author} {\bibfnamefont {R.~A.}\ \bibnamefont
  {Kidd}}, \bibinfo {author} {\bibfnamefont {A.}~\bibnamefont {Safavi-Naini}},\
  and\ \bibinfo {author} {\bibfnamefont {J.~F.}\ \bibnamefont {Corney}},\
  }\href {https://doi.org/10.1103/PhysRevA.102.023330} {\bibfield  {journal}
  {\bibinfo  {journal} {Phys. Rev. A}\ }\textbf {\bibinfo {volume} {102}},\
  \bibinfo {pages} {023330} (\bibinfo {year} {2020})}\BibitemShut {NoStop}%
\bibitem [{\citenamefont {Fortes}\ \emph {et~al.}(2019)\citenamefont {Fortes},
  \citenamefont {Garc{\'{i}}a-Mata}, \citenamefont {Jalabert},\ and\
  \citenamefont {Wisniacki}}]{Fortes2019}%
  \BibitemOpen
  \bibfield  {author} {\bibinfo {author} {\bibfnamefont {E.~M.}\ \bibnamefont
  {Fortes}}, \bibinfo {author} {\bibfnamefont {I.}~\bibnamefont
  {Garc{\'{i}}a-Mata}}, \bibinfo {author} {\bibfnamefont {R.~A.}\ \bibnamefont
  {Jalabert}},\ and\ \bibinfo {author} {\bibfnamefont {D.~A.}\ \bibnamefont
  {Wisniacki}},\ }\href {https://doi.org/10.1103/PhysRevE.100.042201}
  {\bibfield  {journal} {\bibinfo  {journal} {Phys. Rev. E}\ }\textbf {\bibinfo
  {volume} {100}},\ \bibinfo {pages} {042201} (\bibinfo {year}
  {2019})}\BibitemShut {NoStop}%
\bibitem [{\citenamefont {Fortes}\ \emph {et~al.}(2020)\citenamefont {Fortes},
  \citenamefont {Garc{\'{i}}a-Mata}, \citenamefont {Jalabert},\ and\
  \citenamefont {Wisniacki}}]{Fortes2020}%
  \BibitemOpen
  \bibfield  {author} {\bibinfo {author} {\bibfnamefont {E.~M.}\ \bibnamefont
  {Fortes}}, \bibinfo {author} {\bibfnamefont {I.}~\bibnamefont
  {Garc{\'{i}}a-Mata}}, \bibinfo {author} {\bibfnamefont {R.~A.}\ \bibnamefont
  {Jalabert}},\ and\ \bibinfo {author} {\bibfnamefont {D.~A.}\ \bibnamefont
  {Wisniacki}},\ }\href@noop {} {\bibfield  {journal} {\bibinfo  {journal}
  {EPL}\ }\textbf {\bibinfo {volume} {130}},\ \bibinfo {pages} {60001}
  (\bibinfo {year} {2020})}\BibitemShut {NoStop}%
\bibitem [{\citenamefont {D'Alessio}\ \emph {et~al.}(2016)\citenamefont
  {D'Alessio}, \citenamefont {Kafri}, \citenamefont {Polkovnikov},\ and\
  \citenamefont {Rigol}}]{DAlessio2016}%
  \BibitemOpen
  \bibfield  {author} {\bibinfo {author} {\bibfnamefont {L.}~\bibnamefont
  {D'Alessio}}, \bibinfo {author} {\bibfnamefont {Y.}~\bibnamefont {Kafri}},
  \bibinfo {author} {\bibfnamefont {A.}~\bibnamefont {Polkovnikov}},\ and\
  \bibinfo {author} {\bibfnamefont {M.}~\bibnamefont {Rigol}},\ }\href
  {https://doi.org/10.1080/00018732.2016.1198134} {\bibfield  {journal}
  {\bibinfo  {journal} {Adv. Phys.}\ }\textbf {\bibinfo {volume} {65}},\
  \bibinfo {pages} {239} (\bibinfo {year} {2016})}\BibitemShut {NoStop}%
\bibitem [{\citenamefont {Pappalardi}\ \emph {et~al.}(2018)\citenamefont
  {Pappalardi}, \citenamefont {Russomanno}, \citenamefont
  {{\v{Z}}unkovi{\v{c}}}, \citenamefont {Iemini}, \citenamefont {Silva},\ and\
  \citenamefont {Fazio}}]{Pappalardi2018}%
  \BibitemOpen
  \bibfield  {author} {\bibinfo {author} {\bibfnamefont {S.}~\bibnamefont
  {Pappalardi}}, \bibinfo {author} {\bibfnamefont {A.}~\bibnamefont
  {Russomanno}}, \bibinfo {author} {\bibfnamefont {B.}~\bibnamefont
  {{\v{Z}}unkovi{\v{c}}}}, \bibinfo {author} {\bibfnamefont {F.}~\bibnamefont
  {Iemini}}, \bibinfo {author} {\bibfnamefont {A.}~\bibnamefont {Silva}},\ and\
  \bibinfo {author} {\bibfnamefont {R.}~\bibnamefont {Fazio}},\ }\href
  {https://doi.org/10.1103/PhysRevB.98.134303} {\bibfield  {journal} {\bibinfo
  {journal} {Phys. Rev. B}\ }\textbf {\bibinfo {volume} {98}},\ \bibinfo
  {pages} {134303} (\bibinfo {year} {2018})}\BibitemShut {NoStop}%
\bibitem [{\citenamefont {Hummel}\ \emph {et~al.}(2019)\citenamefont {Hummel},
  \citenamefont {Geiger}, \citenamefont {Urbina},\ and\ \citenamefont
  {Richter}}]{Hummel2019}%
  \BibitemOpen
  \bibfield  {author} {\bibinfo {author} {\bibfnamefont {Q.}~\bibnamefont
  {Hummel}}, \bibinfo {author} {\bibfnamefont {B.}~\bibnamefont {Geiger}},
  \bibinfo {author} {\bibfnamefont {J.~D.}\ \bibnamefont {Urbina}},\ and\
  \bibinfo {author} {\bibfnamefont {K.}~\bibnamefont {Richter}},\ }\href
  {https://doi.org/10.1103/PhysRevLett.123.160401} {\bibfield  {journal}
  {\bibinfo  {journal} {Phys. Rev. Lett.}\ }\textbf {\bibinfo {volume} {123}},\
  \bibinfo {pages} {160401} (\bibinfo {year} {2019})}\BibitemShut {NoStop}%
\bibitem [{\citenamefont {Bhattacharyya}\ \emph {et~al.}(2019)\citenamefont
  {Bhattacharyya}, \citenamefont {Chemissany}, \citenamefont {Haque},\ and\
  \citenamefont {Yan}}]{Bhattacharyya2019}%
  \BibitemOpen
  \bibfield  {author} {\bibinfo {author} {\bibfnamefont {A.}~\bibnamefont
  {Bhattacharyya}}, \bibinfo {author} {\bibfnamefont {W.}~\bibnamefont
  {Chemissany}}, \bibinfo {author} {\bibfnamefont {S.~S.}\ \bibnamefont
  {Haque}},\ and\ \bibinfo {author} {\bibfnamefont {B.}~\bibnamefont {Yan}},\
  }\Eprint {https://arxiv.org/abs/1909.01894} {arXiv:1909.01894}  (\bibinfo
  {year} {2019})\BibitemShut {NoStop}%
\bibitem [{\citenamefont {Ali}\ \emph {et~al.}(2020)\citenamefont {Ali},
  \citenamefont {Bhattacharyya}, \citenamefont {Haque}, \citenamefont {Kim},
  \citenamefont {Moynihan},\ and\ \citenamefont {Murugan}}]{Ali2020}%
  \BibitemOpen
  \bibfield  {author} {\bibinfo {author} {\bibfnamefont {T.}~\bibnamefont
  {Ali}}, \bibinfo {author} {\bibfnamefont {A.}~\bibnamefont {Bhattacharyya}},
  \bibinfo {author} {\bibfnamefont {S.~S.}\ \bibnamefont {Haque}}, \bibinfo
  {author} {\bibfnamefont {E.~H.}\ \bibnamefont {Kim}}, \bibinfo {author}
  {\bibfnamefont {N.}~\bibnamefont {Moynihan}},\ and\ \bibinfo {author}
  {\bibfnamefont {J.}~\bibnamefont {Murugan}},\ }\href
  {https://doi.org/10.1103/PhysRevD.101.026021} {\bibfield  {journal} {\bibinfo
   {journal} {Phys. Rev. D}\ }\textbf {\bibinfo {volume} {101}},\ \bibinfo
  {pages} {026021} (\bibinfo {year} {2020})}\BibitemShut {NoStop}%
\bibitem [{\citenamefont {Hashimoto}\ \emph {et~al.}(2020)\citenamefont
  {Hashimoto}, \citenamefont {Huh}, \citenamefont {Kim},\ and\ \citenamefont
  {Watanabe}}]{Hashimoto2020}%
  \BibitemOpen
  \bibfield  {author} {\bibinfo {author} {\bibfnamefont {K.}~\bibnamefont
  {Hashimoto}}, \bibinfo {author} {\bibfnamefont {K.-B.}\ \bibnamefont {Huh}},
  \bibinfo {author} {\bibfnamefont {K.-Y.}\ \bibnamefont {Kim}},\ and\ \bibinfo
  {author} {\bibfnamefont {R.}~\bibnamefont {Watanabe}},\ }\Eprint
  {https://arxiv.org/abs/2007.04746} {arXiv:2007.04746}  (\bibinfo {year}
  {2020})\BibitemShut {NoStop}%
\bibitem [{\citenamefont {Bhattacharyya}\ \emph {et~al.}(2021)\citenamefont
  {Bhattacharyya}, \citenamefont {Chemissany}, \citenamefont {Haque},
  \citenamefont {Murugan},\ and\ \citenamefont {Yan}}]{Bhattacharyya2021}%
  \BibitemOpen
  \bibfield  {author} {\bibinfo {author} {\bibfnamefont {A.}~\bibnamefont
  {Bhattacharyya}}, \bibinfo {author} {\bibfnamefont {W.}~\bibnamefont
  {Chemissany}}, \bibinfo {author} {\bibfnamefont {S.~S.}\ \bibnamefont
  {Haque}}, \bibinfo {author} {\bibfnamefont {J.}~\bibnamefont {Murugan}},\
  and\ \bibinfo {author} {\bibfnamefont {B.}~\bibnamefont {Yan}},\ }\href
  {https://doi.org/10.21468/SciPostPhysCore.4.1.002} {\bibfield  {journal}
  {\bibinfo  {journal} {SciPost Phys. Core}\ }\textbf {\bibinfo {volume} {4}},\
  \bibinfo {pages} {2} (\bibinfo {year} {2021})}\BibitemShut {NoStop}%
\bibitem [{\citenamefont {Dicke}(1954)}]{Dicke1954}%
  \BibitemOpen
  \bibfield  {author} {\bibinfo {author} {\bibfnamefont {R.~H.}\ \bibnamefont
  {Dicke}},\ }\href {https://doi.org/10.1103/PhysRev.93.99} {\bibfield
  {journal} {\bibinfo  {journal} {Phys. Rev.}\ }\textbf {\bibinfo {volume}
  {93}},\ \bibinfo {pages} {99} (\bibinfo {year} {1954})}\BibitemShut {NoStop}%
\bibitem [{\citenamefont {Milburn}\ \emph
  {et~al.}(1997{\natexlab{a}})\citenamefont {Milburn}, \citenamefont {Corney},
  \citenamefont {Wright},\ and\ \citenamefont {Walls}}]{Milburn1997}%
  \BibitemOpen
  \bibfield  {author} {\bibinfo {author} {\bibfnamefont {G.~J.}\ \bibnamefont
  {Milburn}}, \bibinfo {author} {\bibfnamefont {J.}~\bibnamefont {Corney}},
  \bibinfo {author} {\bibfnamefont {E.~M.}\ \bibnamefont {Wright}},\ and\
  \bibinfo {author} {\bibfnamefont {D.~F.}\ \bibnamefont {Walls}},\ }\href
  {https://doi.org/10.1103/PhysRevA.55.4318} {\bibfield  {journal} {\bibinfo
  {journal} {Phys. Rev. A}\ }\textbf {\bibinfo {volume} {55}},\ \bibinfo
  {pages} {4318} (\bibinfo {year} {1997}{\natexlab{a}})}\BibitemShut {NoStop}%
\bibitem [{\citenamefont {Milburn}\ \emph
  {et~al.}(1997{\natexlab{b}})\citenamefont {Milburn}, \citenamefont {Corney},
  \citenamefont {Harris}, \citenamefont {Wright},\ and\ \citenamefont
  {Walls}}]{Milburn1997chaos}%
  \BibitemOpen
  \bibfield  {author} {\bibinfo {author} {\bibfnamefont {G.~J.}\ \bibnamefont
  {Milburn}}, \bibinfo {author} {\bibfnamefont {J.~F.}\ \bibnamefont {Corney}},
  \bibinfo {author} {\bibfnamefont {D.}~\bibnamefont {Harris}}, \bibinfo
  {author} {\bibfnamefont {E.~M.}\ \bibnamefont {Wright}},\ and\ \bibinfo
  {author} {\bibfnamefont {D.~F.}\ \bibnamefont {Walls}},\ }in\ \href
  {https://doi.org/10.1117/12.273762} {\emph {\bibinfo {booktitle} {Photonics
  West ’97: Atom Optics}}},\ \bibinfo {series} {SPIE Proceedings}, Vol.\
  \bibinfo {volume} {2995},\ \bibinfo {editor} {edited by\ \bibinfo {editor}
  {\bibfnamefont {M.~G.}\ \bibnamefont {Prentiss}}\ and\ \bibinfo {editor}
  {\bibfnamefont {W.~D.}\ \bibnamefont {Phillips}}}\ (\bibinfo  {publisher}
  {SPIE},\ \bibinfo {address} {Bellingham, WA},\ \bibinfo {year} {1997})\ pp.\
  \bibinfo {pages} {232--239}\BibitemShut {NoStop}%
\bibitem [{\citenamefont {Kidd}\ \emph {et~al.}(2019)\citenamefont {Kidd},
  \citenamefont {Olsen},\ and\ \citenamefont {Corney}}]{Kidd2019}%
  \BibitemOpen
  \bibfield  {author} {\bibinfo {author} {\bibfnamefont {R.~A.}\ \bibnamefont
  {Kidd}}, \bibinfo {author} {\bibfnamefont {M.~K.}\ \bibnamefont {Olsen}},\
  and\ \bibinfo {author} {\bibfnamefont {J.~F.}\ \bibnamefont {Corney}},\
  }\href {https://doi.org/10.1103/PhysRevA.100.013625} {\bibfield  {journal}
  {\bibinfo  {journal} {Phys. Rev. A}\ }\textbf {\bibinfo {volume} {100}},\
  \bibinfo {pages} {013625} (\bibinfo {year} {2019})}\BibitemShut {NoStop}%
\bibitem [{\citenamefont {Ch{\'{a}}vez-Carlos}\ \emph
  {et~al.}(2016)\citenamefont {Ch{\'{a}}vez-Carlos}, \citenamefont
  {Bastarrachea-Magnani}, \citenamefont {Lerma-Hern{\'{a}}ndez},\ and\
  \citenamefont {Hirsch}}]{Chavez-Carlos2016}%
  \BibitemOpen
  \bibfield  {author} {\bibinfo {author} {\bibfnamefont {J.}~\bibnamefont
  {Ch{\'{a}}vez-Carlos}}, \bibinfo {author} {\bibfnamefont {M.~A.}\
  \bibnamefont {Bastarrachea-Magnani}}, \bibinfo {author} {\bibfnamefont
  {S.}~\bibnamefont {Lerma-Hern{\'{a}}ndez}},\ and\ \bibinfo {author}
  {\bibfnamefont {J.~G.}\ \bibnamefont {Hirsch}},\ }\href
  {https://doi.org/10.1103/PhysRevE.94.022209} {\bibfield  {journal} {\bibinfo
  {journal} {Phys. Rev. E}\ }\textbf {\bibinfo {volume} {94}},\ \bibinfo
  {pages} {022209} (\bibinfo {year} {2016})}\BibitemShut {NoStop}%
\bibitem [{\citenamefont {Bastarrachea-Magnani}\ \emph
  {et~al.}(2017)\citenamefont {Bastarrachea-Magnani}, \citenamefont
  {{L{\'{o}}pez-del-Carpio}}, \citenamefont {Ch{\'{a}}vez-Carlos},
  \citenamefont {Lerma-Hern{\'{a}}ndez},\ and\ \citenamefont
  {Hirsch}}]{Bastarrachea-Magnani2017}%
  \BibitemOpen
  \bibfield  {author} {\bibinfo {author} {\bibfnamefont {M.~A.}\ \bibnamefont
  {Bastarrachea-Magnani}}, \bibinfo {author} {\bibfnamefont {B.}~\bibnamefont
  {{L{\'{o}}pez-del-Carpio}}}, \bibinfo {author} {\bibfnamefont
  {J.}~\bibnamefont {Ch{\'{a}}vez-Carlos}}, \bibinfo {author} {\bibfnamefont
  {S.}~\bibnamefont {Lerma-Hern{\'{a}}ndez}},\ and\ \bibinfo {author}
  {\bibfnamefont {J.~G.}\ \bibnamefont {Hirsch}},\ }\href
  {https://doi.org/10.1088/1402-4896/aa6640} {\bibfield  {journal} {\bibinfo
  {journal} {Phys. Scr.}\ }\textbf {\bibinfo {volume} {92}},\ \bibinfo {pages}
  {054003} (\bibinfo {year} {2017})}\BibitemShut {NoStop}%
\bibitem [{\citenamefont {Ch{\'{a}}vez-Carlos}\ \emph
  {et~al.}(2019)\citenamefont {Ch{\'{a}}vez-Carlos}, \citenamefont
  {{L{\'{o}}pez-del-Carpio}}, \citenamefont {Bastarrachea-Magnani},
  \citenamefont {Str{\'{a}}nsk{\'{y}}}, \citenamefont {Lerma-Hern{\'{a}}ndez},
  \citenamefont {Santos},\ and\ \citenamefont {Hirsch}}]{Chavez-Carlos2019}%
  \BibitemOpen
  \bibfield  {author} {\bibinfo {author} {\bibfnamefont {J.}~\bibnamefont
  {Ch{\'{a}}vez-Carlos}}, \bibinfo {author} {\bibfnamefont {B.}~\bibnamefont
  {{L{\'{o}}pez-del-Carpio}}}, \bibinfo {author} {\bibfnamefont {M.~A.}\
  \bibnamefont {Bastarrachea-Magnani}}, \bibinfo {author} {\bibfnamefont
  {P.}~\bibnamefont {Str{\'{a}}nsk{\'{y}}}}, \bibinfo {author} {\bibfnamefont
  {S.}~\bibnamefont {Lerma-Hern{\'{a}}ndez}}, \bibinfo {author} {\bibfnamefont
  {L.~F.}\ \bibnamefont {Santos}},\ and\ \bibinfo {author} {\bibfnamefont
  {J.~G.}\ \bibnamefont {Hirsch}},\ }\href
  {https://doi.org/10.1103/PhysRevLett.122.024101} {\bibfield  {journal}
  {\bibinfo  {journal} {Phys. Rev. Lett.}\ }\textbf {\bibinfo {volume} {122}},\
  \bibinfo {pages} {024101} (\bibinfo {year} {2019})}\BibitemShut {NoStop}%
\bibitem [{\citenamefont {Bastarrachea-Magnani}\ \emph
  {et~al.}(2016)\citenamefont {Bastarrachea-Magnani}, \citenamefont
  {{{L{\'{o}}pez-del-Carpio}}}, \citenamefont {Ch{\'{a}}vez-Carlos},
  \citenamefont {Lerma-Hern{\'{a}}ndez},\ and\ \citenamefont
  {Hirsch}}]{Bastarrachea-Magnani2016}%
  \BibitemOpen
  \bibfield  {author} {\bibinfo {author} {\bibfnamefont {M.~A.}\ \bibnamefont
  {Bastarrachea-Magnani}}, \bibinfo {author} {\bibfnamefont {B.}~\bibnamefont
  {{{L{\'{o}}pez-del-Carpio}}}}, \bibinfo {author} {\bibfnamefont
  {J.}~\bibnamefont {Ch{\'{a}}vez-Carlos}}, \bibinfo {author} {\bibfnamefont
  {S.}~\bibnamefont {Lerma-Hern{\'{a}}ndez}},\ and\ \bibinfo {author}
  {\bibfnamefont {J.~G.}\ \bibnamefont {Hirsch}},\ }\href
  {https://doi.org/10.1103/PhysRevE.93.022215} {\bibfield  {journal} {\bibinfo
  {journal} {Phys. Rev. E}\ }\textbf {\bibinfo {volume} {93}},\ \bibinfo
  {pages} {022215} (\bibinfo {year} {2016})}\BibitemShut {NoStop}%
\bibitem [{\citenamefont {Chen}\ \emph {et~al.}(2008)\citenamefont {Chen},
  \citenamefont {Zhang}, \citenamefont {Liu},\ and\ \citenamefont
  {Wang}}]{ECB0}%
  \BibitemOpen
  \bibfield  {author} {\bibinfo {author} {\bibfnamefont {Q.-H.}\ \bibnamefont
  {Chen}}, \bibinfo {author} {\bibfnamefont {Y.-Y.}\ \bibnamefont {Zhang}},
  \bibinfo {author} {\bibfnamefont {T.}~\bibnamefont {Liu}},\ and\ \bibinfo
  {author} {\bibfnamefont {K.-L.}\ \bibnamefont {Wang}},\ }\href
  {https://doi.org/10.1103/PhysRevA.78.051801} {\bibfield  {journal} {\bibinfo
  {journal} {Phys. Rev. A}\ }\textbf {\bibinfo {volume} {78}},\ \bibinfo
  {pages} {051801(R)} (\bibinfo {year} {2008})}\BibitemShut {NoStop}%
\bibitem [{\citenamefont {Bastarrachea-Magnani}\ and\ \citenamefont
  {Hirsch}(2014)}]{ECB1}%
  \BibitemOpen
  \bibfield  {author} {\bibinfo {author} {\bibfnamefont {M.~A.}\ \bibnamefont
  {Bastarrachea-Magnani}}\ and\ \bibinfo {author} {\bibfnamefont {J.~G.}\
  \bibnamefont {Hirsch}},\ }\href
  {https://doi.org/10.1088/0031-8949/2014/T160/014005} {\bibfield  {journal}
  {\bibinfo  {journal} {Phys. Scr.}\ }\textbf {\bibinfo {volume} {T160}},\
  \bibinfo {pages} {014005} (\bibinfo {year} {2014})}\BibitemShut {NoStop}%
\bibitem [{\citenamefont {Hirsch}\ and\ \citenamefont
  {Bastarrachea-Magnani}(2014)}]{ECB2}%
  \BibitemOpen
  \bibfield  {author} {\bibinfo {author} {\bibfnamefont {J.~G.}\ \bibnamefont
  {Hirsch}}\ and\ \bibinfo {author} {\bibfnamefont {M.~A.}\ \bibnamefont
  {Bastarrachea-Magnani}},\ }\href
  {https://doi.org/10.1088/0031-8949/2014/T160/014018} {\bibfield  {journal}
  {\bibinfo  {journal} {Phys. Scr.}\ }\textbf {\bibinfo {volume} {T160}},\
  \bibinfo {pages} {014018} (\bibinfo {year} {2014})}\BibitemShut {NoStop}%
\bibitem [{\citenamefont {Arecchi}\ \emph {et~al.}(1972)\citenamefont
  {Arecchi}, \citenamefont {Courtens}, \citenamefont {Gilmore},\ and\
  \citenamefont {Thomas}}]{Arecchi1972}%
  \BibitemOpen
  \bibfield  {author} {\bibinfo {author} {\bibfnamefont {F.~T.}\ \bibnamefont
  {Arecchi}}, \bibinfo {author} {\bibfnamefont {E.}~\bibnamefont {Courtens}},
  \bibinfo {author} {\bibfnamefont {R.}~\bibnamefont {Gilmore}},\ and\ \bibinfo
  {author} {\bibfnamefont {H.}~\bibnamefont {Thomas}},\ }\href
  {https://doi.org/10.1103/PhysRevA.6.2211} {\bibfield  {journal} {\bibinfo
  {journal} {Phys. Rev. A}\ }\textbf {\bibinfo {volume} {6}},\ \bibinfo {pages}
  {2211} (\bibinfo {year} {1972})}\BibitemShut {NoStop}%
\bibitem [{\citenamefont {Schmitt}\ \emph {et~al.}(2019)\citenamefont
  {Schmitt}, \citenamefont {Sels}, \citenamefont {Kehrein},\ and\ \citenamefont
  {Polkovnikov}}]{Schmitt2019}%
  \BibitemOpen
  \bibfield  {author} {\bibinfo {author} {\bibfnamefont {M.}~\bibnamefont
  {Schmitt}}, \bibinfo {author} {\bibfnamefont {D.}~\bibnamefont {Sels}},
  \bibinfo {author} {\bibfnamefont {S.}~\bibnamefont {Kehrein}},\ and\ \bibinfo
  {author} {\bibfnamefont {A.}~\bibnamefont {Polkovnikov}},\ }\href
  {https://doi.org/10.1103/PhysRevB.99.134301} {\bibfield  {journal} {\bibinfo
  {journal} {Phys. Rev. B}\ }\textbf {\bibinfo {volume} {99}},\ \bibinfo
  {pages} {134301} (\bibinfo {year} {2019})}\BibitemShut {NoStop}%
\bibitem [{Note1()}]{Note1}%
  \BibitemOpen
  \bibinfo {note} {That the FOTOC corresponds to a variance, $\protect \var
  {(\protect \hat {S}_x)}$ gives rise to one factor of two~\cite
  {Lewis-Swan2019}. Additionally, the initial phase-space trajectories are
  orthogonal to the $ \langle \protect \hat {S}_x\rangle $-axis. The change in
  $ \langle \protect \hat {S}_x\rangle $ is thus a second-order effect, which
  gives rise to a second factor of two.}\BibitemShut {Stop}%
\bibitem [{\citenamefont {Skokos}(2010)}]{Skokos2010}%
  \BibitemOpen
  \bibfield  {author} {\bibinfo {author} {\bibfnamefont {C.}~\bibnamefont
  {Skokos}},\ }in\ \href {https://doi.org/10.1007/978-3-642-04458-8_2} {\emph
  {\bibinfo {booktitle} {Dynamics of Small Solar System Bodies and
  Exoplanets}}},\ \bibinfo {series} {Lect. Notes Phys.}, Vol.\ \bibinfo
  {volume} {790},\ \bibinfo {editor} {edited by\ \bibinfo {editor}
  {\bibfnamefont {J.~J.}\ \bibnamefont {Souchay}}\ and\ \bibinfo {editor}
  {\bibfnamefont {R.}~\bibnamefont {Dvorak}}}\ (\bibinfo  {publisher}
  {Springer},\ \bibinfo {address} {Berlin, Heidelberg},\ \bibinfo {year}
  {2010})\ pp.\ \bibinfo {pages} {63--135}\BibitemShut {NoStop}%
\bibitem [{\citenamefont {Zeni}\ and\ \citenamefont {Gallas}(1995)}]{Zeni1995}%
  \BibitemOpen
  \bibfield  {author} {\bibinfo {author} {\bibfnamefont {A.~R.}\ \bibnamefont
  {Zeni}}\ and\ \bibinfo {author} {\bibfnamefont {J.~A.}\ \bibnamefont
  {Gallas}},\ }\href {https://doi.org/10.1016/0167-2789(95)00215-4} {\bibfield
  {journal} {\bibinfo  {journal} {Physica D}\ }\textbf {\bibinfo {volume}
  {89}},\ \bibinfo {pages} {71} (\bibinfo {year} {1995})}\BibitemShut {NoStop}%
\bibitem [{Note2()}]{Note2}%
  \BibitemOpen
  \bibinfo {note} {We note some regularly-spaced `pulses' in the OTOC over
  intermediate time-scales. The timing of this transient feature seems to
  correlate with the driving period.}\BibitemShut {Stop}%
\bibitem [{\citenamefont {Rigol}\ \emph {et~al.}(2008)\citenamefont {Rigol},
  \citenamefont {Dunjko},\ and\ \citenamefont {Olshanii}}]{Rigol2008}%
  \BibitemOpen
  \bibfield  {author} {\bibinfo {author} {\bibfnamefont {M.}~\bibnamefont
  {Rigol}}, \bibinfo {author} {\bibfnamefont {V.}~\bibnamefont {Dunjko}},\ and\
  \bibinfo {author} {\bibfnamefont {M.}~\bibnamefont {Olshanii}},\ }\href
  {https://doi.org/10.1038/nature06838} {\bibfield  {journal} {\bibinfo
  {journal} {Nature}\ }\textbf {\bibinfo {volume} {452}},\ \bibinfo {pages}
  {854} (\bibinfo {year} {2008})}\BibitemShut {NoStop}%
\bibitem [{\citenamefont {Fr{\"{o}}wis}\ \emph {et~al.}(2015)\citenamefont
  {Fr{\"{o}}wis}, \citenamefont {Schmied},\ and\ \citenamefont
  {Gisin}}]{Frowis2015}%
  \BibitemOpen
  \bibfield  {author} {\bibinfo {author} {\bibfnamefont {F.}~\bibnamefont
  {Fr{\"{o}}wis}}, \bibinfo {author} {\bibfnamefont {R.}~\bibnamefont
  {Schmied}},\ and\ \bibinfo {author} {\bibfnamefont {N.}~\bibnamefont
  {Gisin}},\ }\href {https://doi.org/10.1103/PhysRevA.92.012102} {\bibfield
  {journal} {\bibinfo  {journal} {Phys. Rev. A}\ }\textbf {\bibinfo {volume}
  {92}},\ \bibinfo {pages} {012102} (\bibinfo {year} {2015})}\BibitemShut
  {NoStop}%
\bibitem [{\citenamefont {Bianchi}\ and\ \citenamefont
  {Don{\`{a}}}(2019)}]{Bianchi2019}%
  \BibitemOpen
  \bibfield  {author} {\bibinfo {author} {\bibfnamefont {E.}~\bibnamefont
  {Bianchi}}\ and\ \bibinfo {author} {\bibfnamefont {P.}~\bibnamefont
  {Don{\`{a}}}},\ }\href {https://doi.org/10.1103/PhysRevD.100.105010}
  {\bibfield  {journal} {\bibinfo  {journal} {Phys. Rev. D}\ }\textbf {\bibinfo
  {volume} {100}},\ \bibinfo {pages} {105010} (\bibinfo {year}
  {2019})}\BibitemShut {NoStop}%
\bibitem [{\citenamefont {D'Alessio}\ and\ \citenamefont
  {Rigol}(2014)}]{DAlessio2014}%
  \BibitemOpen
  \bibfield  {author} {\bibinfo {author} {\bibfnamefont {L.}~\bibnamefont
  {D'Alessio}}\ and\ \bibinfo {author} {\bibfnamefont {M.}~\bibnamefont
  {Rigol}},\ }\href {https://doi.org/10.1103/PhysRevX.4.041048} {\bibfield
  {journal} {\bibinfo  {journal} {Phys. Rev. X}\ }\textbf {\bibinfo {volume}
  {4}},\ \bibinfo {pages} {041048} (\bibinfo {year} {2014})}\BibitemShut
  {NoStop}%
\bibitem [{\citenamefont {Walls}\ and\ \citenamefont
  {Milburn}(2008)}]{Walls2008}%
  \BibitemOpen
  \bibfield  {author} {\bibinfo {author} {\bibfnamefont {D.}~\bibnamefont
  {Walls}}\ and\ \bibinfo {author} {\bibfnamefont {G.~J.}\ \bibnamefont
  {Milburn}},\ }\href {https://doi.org/10.1007/978-3-540-28574-8} {\emph
  {\bibinfo {title} {{Quantum Optics}}}},\ \bibinfo {edition} {2nd}\ ed.\
  (\bibinfo  {publisher} {Springer},\ \bibinfo {address} {Berlin, Heidelberg},\
  \bibinfo {year} {2008})\BibitemShut {NoStop}%
\bibitem [{\citenamefont {Olsen}\ and\ \citenamefont
  {Bradley}(2009)}]{Olsen2009}%
  \BibitemOpen
  \bibfield  {author} {\bibinfo {author} {\bibfnamefont {M.}~\bibnamefont
  {Olsen}}\ and\ \bibinfo {author} {\bibfnamefont {A.}~\bibnamefont
  {Bradley}},\ }\href {https://doi.org/10.1016/j.optcom.2009.06.033} {\bibfield
   {journal} {\bibinfo  {journal} {Opt. Commun.}\ }\textbf {\bibinfo {volume}
  {282}},\ \bibinfo {pages} {3924} (\bibinfo {year} {2009})}\BibitemShut
  {NoStop}%
\bibitem [{\citenamefont {Polkovnikov}(2010)}]{Polkovnikov2010}%
  \BibitemOpen
  \bibfield  {author} {\bibinfo {author} {\bibfnamefont {A.}~\bibnamefont
  {Polkovnikov}},\ }\href {https://doi.org/10.1016/j.aop.2010.02.006}
  {\bibfield  {journal} {\bibinfo  {journal} {Ann. Phys. (N. Y.)}\ }\textbf
  {\bibinfo {volume} {325}},\ \bibinfo {pages} {1790} (\bibinfo {year}
  {2010})}\BibitemShut {NoStop}%
\bibitem [{\citenamefont {Ruostekoski}\ and\ \citenamefont
  {Martin}(2013)}]{Ruostekoski2013}%
  \BibitemOpen
  \bibfield  {author} {\bibinfo {author} {\bibfnamefont {J.}~\bibnamefont
  {Ruostekoski}}\ and\ \bibinfo {author} {\bibfnamefont {A.~D.}\ \bibnamefont
  {Martin}},\ }in\ \href {https://doi.org/10.1142/9781848168121_0013} {\emph
  {\bibinfo {booktitle} {Quantum Gases}}},\ \bibinfo {series} {Cold Atoms},
  Vol.~\bibinfo {volume} {1},\ \bibinfo {editor} {edited by\ \bibinfo {editor}
  {\bibfnamefont {N.}~\bibnamefont {Proukakis}}, \bibinfo {editor}
  {\bibfnamefont {S.}~\bibnamefont {Gardiner}}, \bibinfo {editor}
  {\bibfnamefont {M.}~\bibnamefont {Davis}},\ and\ \bibinfo {editor}
  {\bibfnamefont {M.}~\bibnamefont {Szyma{\'n}ska}}}\ (\bibinfo  {publisher}
  {Imperial College Press},\ \bibinfo {address} {London},\ \bibinfo {year}
  {2013})\ Chap.~\bibinfo {chapter} {13}, pp.\ \bibinfo {pages}
  {203--214}\BibitemShut {NoStop}%
\bibitem [{Note3()}]{Note3}%
  \BibitemOpen
  \bibinfo {note} {We quote unnormalised values since there is no upper bound
  to the diagonal entropy in the infinite-dimensional Hilbert space of the
  Dicke model.}\BibitemShut {Stop}%
\end{thebibliography}


\end{document}